\newif\ifPNASstyle
\PNASstyletrue
\PNASstylefalse

\ifPNASstyle
   \documentclass[final,titlepage]{pnastwo}
   \usepackage{t1enc}
   \usepackage{PNAStwoF}
\else
   \documentclass[10pt,english]{article}
\fi

\usepackage{bm}
\usepackage{amsmath}
\usepackage{amssymb}
\usepackage{amstext}
\usepackage{graphicx}
\usepackage{cite}
\usepackage{color} 
\usepackage[english]{babel}

\ifPNASstyle
   \contributor{Submitted to Proceedings of the National Academy of Sciences of the United States of America}
   \url{}
   \copyrightyear{2013}
   \issuedate{Issue Date}
   \volume{Volume}
   \issuenumber{Issue Number}
\else
   \usepackage{multirow}
   \usepackage[labelfont=bf,labelsep=period,justification=raggedright]{caption}
   \makeatletter
   \renewcommand{\@biblabel}[1]{\quad#1.}
   \makeatother
\fi

\ifPNASstyle
\else
   \date{}
\fi

\begin{document}

\ifPNASstyle
   \title{Genotype to phenotype mapping and
     the fitness landscape of the \emph{E.~coli lac} promoter}

   \author{%
     Jakub Otwinowski%
     \thanks{E-mail:jotwinowski@physics.emory.edu}%
     \affil{1}{Department of Physics, Emory University, Atlanta, GA 30322, USA}%
     \affil{2}{Department of Biology, University of Pennsylvania, Philadelphia, PA 19104, USA}, %
     \and Ilya Nemenman%
     \thanks{E-mail: ilya.nemenman@emory.edu}%
     \affil{3}{Department of Physics, Department of Biology, and %
       Computational and Life Sciences Initiative, Emory University, Atlanta, GA 30322, USA}
   }
   \maketitle
   \begin{article}

\else
   \begin{flushleft} {\Large \textbf{Genotype to phenotype mapping and
         the fitness landscape of the \emph{E.~coli lac} promoter} }
     \\
     Jakub Otwinowski$^{1,2}$, 
     Ilya Nemenman$^{3}$, 
     \\
     {\bf 1} Department of Physics, Emory University, Atlanta, GA 30322, USA
     \\
     {\bf 2} Department of Biology, University of Pennsylvania, Philadelphia, PA 19104, USA
     \\
     {\bf 3} Department of Physics, Department of Biology, and Computational
     and Life Sciences Initiative, Emory University, Atlanta, GA 30322, USA
     \\
     $\ast$ E-mail: jotwinowski@physics.emory.edu, ilya.nemenman@emory.edu
   \end{flushleft}
\fi

\ifPNASstyle
   \begin{abstract}
\else
   \section*{Abstract}
\fi

Genotype-to-phenotype maps and the related fitness landscapes that
include epistatic interactions are difficult to measure because of
their high dimensional structure. Here we construct such a map using
the recently collected corpora of high-throughput sequence data from
the 75 base pairs long mutagenized \emph{E. coli lac} promoter region,
where each sequence is associated with its phenotype, the induced
transcriptional activity measured by a fluorescent reporter. We find
that the additive (non-epistatic) contributions of individual
mutations account for about two-thirds of the explainable phenotype
variance, while pairwise epistasis explains about 7\% of the variance
for the full mutagenized sequence and about 15\% for the subsequence
associated with protein binding sites. Surprisingly, there is no
evidence for third order epistatic contributions, and our inferred
fitness landscape is essentially single peaked, with a small amount of
antagonistic epistasis. There is a significant selective pressure on
the wild type, which we deduce to be multi-objective optimal for gene
expression in environments with different nutrient sources. We
identify transcription factor (CRP) and RNA polymerase binding sites
in the promotor region and their interactions without difficult
optimization steps.  In particular, we observe evidence for previously
unexplored genetic regulatory mechanisms, possibly kinetic in
nature. We conclude with a cautionary note that inferred properties of
fitness landscapes may be severely influenced by biases in the
sequence data.

\ifPNASstyle
   \end{abstract}
\fi

\section{Introduction}

\ifPNASstyle
   \dropcap{M}any%
\else
   Many%
\fi aspects of evolution, such as selection, recombination, and
speciation, depend on the relationships between genotype, phenotype, and
fitness.  These relationships often involve complex and collective
effects \cite{Goldenfeld2011}, which are difficult to untangle. One
approach is to measure the fitness of many different genotypes, and
build a \emph{fitness landscape}, a high dimensional map from
genotype/phenotype to reproductive fitness.  This concept was first
introduced by Sewell Wright in 1932 \cite{Wright1932}.  Evolutionary
dynamics and adaptation depend crucially on features of the fitness
landscape, and many studies have quantified large scale features of
landscapes, including genetic interactions
\cite{Szendro2012,Chou2011,Franke2011a,Khan2011a,Weinreich:2006ig,Hall2010,DaSilva2010,Lunzer2005},
the presence of stabilizing selection \cite{Kingsolver2001,Shaw2010},
or the reproducibility of evolutionary paths
\cite{Poelwijk2007,Weinreich:2006ig}.  

A major difficulty that has precluded mapping of large fitness
landscape, is \emph{epistasis}, which is the dependence of fitness
effects of a mutation on the presence of other mutations. Epistasis
makes the inference of landscapes combinatorially complex. This
problem has attracted substantial attention. For example, millions of
interactions between gene pairs have been measured from genetic
knockout experiments
\cite{Segre2005,Costanzo2010,Moore2005,Phillips2008,Stark2011,Baryshnikova2010}.
Higher order epistatic interactions, that is those involving more than
two loci at a time, have also been investigated for small fitness
landscapes \cite{Szendro2012}.

Another popular approach is mapping genotypes to phenotypes (also
known as the Quantitative trait loci or QTL analysis \cite{Hu-97}),
which includes the dimensionality reduction problem, but is simpler
since many phenotypes are easier to quantify reliably than the number
of progenies, which exhibits large fluctuations. One then separately
studies the lower dimensional map from the phenotype to the
reproductive rate to complete the construction of the fitness
landscape.

Unfortunately, few of these pioneering studies have provided a
genotype to phenotype or to fitness mapping for longer genetic
sequences, and most such large maps are modeled without epistasis
(see, e.~g.,~\cite{Brem2005}).  Indeed, a complete landscape would be
defined not by genes or specific loci, but by all possible nucleotide
sequences. However with $\sim4^{L}$ different sequences of length $L$,
it had been impractical to measure the landscapes for sequences of
relatively large length until next generation sequencing technologies
dramatically lowered the cost \cite{Shendure2008}. Nonetheless,
measuring phenotypes of a large number of sequences is still tricky,
and only a few large fitness landscapes have been quantified. For
example, Pitt et al.~measured the fitness landscape of $\sim10^{7}$
RNA sequences with an in vitro selection protocol
\cite{Pitt2010}. Similarly, Mora et al.\ studied frequencies of
genetic sequences of IgM molecules in zebrafish B cells (which are
related to fitnesses), but they imposed a translational symmetry of
the sequence \cite{Mora2010}.  Finally, Hinkley et al.~analyzed 70,000
HIV sequences and their \emph{in vitro} fitnesses, built a fitness
landscape defined on different amino acids of certain HIV genes, and
then investigated large scale properties of the ensuing landscape
\cite{Hinkley2011,Kouyos2012}.  However, even in these high throughput
studies, the data did not contain all possible pairs of mutations,
potentially biasing the results, especially far from the wild type
sequences (see {\em Discussion}).

In this article, we reconstruct a large, yet detailed bacterial
genotype to phenotype map, including quantifying the epistatic
interactions in the ensuing fitness landscape. We seek a 
landscape based on long {\em nucleotide} sequences, which additionally
allows quantifying phenotypes of transcriptional regulation in
addition to those of enzymatic activity. This permits fitnesses to be
defined over both coding and non-coding DNA. To map the landscape far
from the wild type genotype, we would like sampling of the sequence
data that is unbiased by selection.

Recent experiments by Kinney et al.~\cite{Kinney2010a} have collected
a dataset that comes close to satisfying these criteria. The data
consists of mutagenized transcriptional regulatory sequences from the
\emph{E.\,coli} (MG1655 and TK310 strains) \emph{lac} promoter. In
total, there were $\sim129,000$ \emph{lac} promoter sequences
mutagenized in a 75 nucleotide region containing the cAMP receptor
protein (CRP) and RNA polymerase (RNAP) binding sites (-75:-1), with
$6.8\pm 2.7$ mutations per sequence (mean $\pm$ standard deviation) (see
Ref.~\cite{Kinney2010a} for additional data set details). The
transcriptional activity induced by the mutagenized promoters was
measured through fluorescence of the transcribed gene products and
FACS sorted according to the transcriptional activity into up to nine
logarithmically spaced categories. All categories were then
independently sequenced, so that the quantitative (on the scale of 1
to 9) phenotypic effect of each sequence is known to within a certain
accuracy. Further, there were an additional $\sim 52,000$
sequence-expression pairs for the same operon analyzes in different
enviornmental conditions.  Thus the data can be used to reconstruct
the genotype-to-phenotype map. However, the promotor activity is
directly related to lactose metabolism and thus is correlated with
growth rate or fitness under conditions where lactose is the preferred
energy source. Therefore, the fluorescence may also be viewed as a
proxy for fitness of this sequence.

In summary, the Kinney et al.~\cite{Kinney2010a} dataset provides
simultaneous measurements of sequences and their phenotype. Crucially,
the data set is dense, so that every pair of mutations has occurred at
least 20 times, each time in a different genetic backgrounds of about
5 other random mutations.  We use these sequence and transcriptional
activity data to infer the detailed genetic landscape for the 75
nucleotide DNA sequence, quantifying pairwise epistatic interactions
among all of the nucleotides to the accuracy afforded by the
data. This is done by constructing a linear-nonlinear regression model
that connects sequences to their phenotypes. Since the number of
possible epistatic interactions is comparable with the number of
sampled sequences, we control the complexity of the models by $L_{1}$
regularization, and hence prevent overfitting. This also imposes
sparsity on the epistatic interactions, which we expect from the
limited number of binding sites.  We then analyze the statistics of
epistatic effects in the inferred landscape. Finally, analysis of the
landscapes obtained under different environmental conditions provides
evidence that the wild-type sequence of the \emph{E.~coli lac}
promoter is close to optimal in the ecological niche that the
bacterium occupies.

\section{Results}
\subsection{Inferring the non-epistatic genotype to phenotype map }

The simplest model of a genotype to phenotype map is one where each
locus contributes a fixed amount to the phenotype, regardless of the
state of other loci. Thus we used the sequence and the fluorescence
measurements (see {\em Methods}) to fit an additive map using linear
regression of the fluorescence values $y$ (integers 1 to 9) on the
genetic code which are treated as 75 categorical variables with four
levels: A,T,G,C. The dummy variables encode the presence of mutations
relative to the wild type ($x_{i}=1$ when a mutation is present, and
$x_{i}=0$ otherwise). Since there are four nucleic acids, each locus
has three binary numbers for each of the possible mutations from the
wild-type, and the sequence length is effectively tripled. In other
words, for each locus, 000 represents the wild-type, and 001, 010, 100
represent the three mutations (see Table \ref{tbl:code} in {\em
  Methods}). The statistical model is
\begin{equation}
  f(y^{(a)})=\beta_{0}+\sum_{j=1}^{3L}\beta_{j}x_{j}^{(a)}+\varepsilon^{(a)},\label{eq:linear}
\end{equation}
where $\varepsilon$ is the statistical noise, and the superscript
$(a)$ stands for a single bacterium, for which the sequence,
$x_j^{(a)}$, and the fluorescence, $y^{(a)}$, are known. In subsequent
equations, the superscript is suppressed for brevity. Part of the
genotype-phenotype map may be non-linear due to the mapping from
fluorescence to bin number and due to some remaining background
fluorescence. Thus we replace $y$ with a non-linear monotonic function
$f(y)$ chosen to optimize the explanatory power of the nonepistatic
statistical model, and likely bias downwards inferred effects of
epistatic contributions (see {\em Methods}). The coefficients,
$\beta_0$ and $\beta_j$, are found by ordinary least squares
regression, e.~g., coefficients that minimize
$\langle\varepsilon^{2}\rangle$ in Eq.~(\ref{eq:linear}). Since the
wild-type is a sequence of all zeros, $\beta_{0}$ is the predicted
phenotype of the wild type.

The coefficient $r^{2}=1-\sigma_{\varepsilon}^{2}/\sigma_{f(y)}^{2}$
measures the goodness of fit, or how much of the variance in the data,
$\sigma^2_y$, is explained by the model. The linear model yields
$r^{2}=0.514\pm0.002$.

Some variation in the data is experimental noise, such as background
fluorescence and cell-to-cell variability, and sets an upper bound on
the possible $r^{2}$. In {\em Methods}, we estimate this {\em
  intrinsic} noise to be 10-24\%, and therefore about 76-90\% of the
total variability of the data can be explained by {\em any}
statistical model, even an arbitrarily complex model. Therefore the
linear model accounts for 57-67\% of the explainable variance.  We
emphasize that this statement is not about mechanistic underpinnings
of the genotype-to-phenotype relation, but about statistics of the
data only. As in any multivariate model, it is possible for the
statistical linear effects to emerge from superposition of many
mechanistic epistatic interactions.

\begin{figure}
\centerline{\includegraphics[width=3.5in]{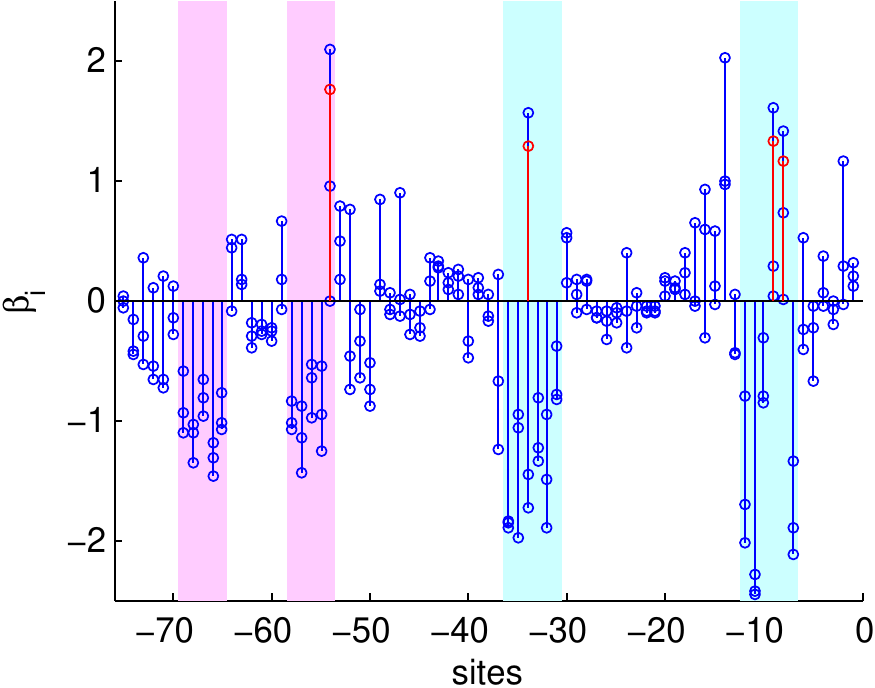}}
\caption{\label{fig:linearcoeff}Stem plot of the linear
  coefficients. Three circles on each stem represent the changes in
  phenotype for each of the three possible mutations per site. CRP and RNAP are known to each bind at two sites (magenta and cyan areas). Red circles correspond to the mutations
  needed to get the consensus sequences.}
\end{figure}

Examination of the coefficients $\beta_{j}$ with the largest magnitude
reveals the consensus locations of the CRP and RNAP binding sites
(Fig.~\ref{fig:linearcoeff}), which validates the modeling approach.
Interestingly, the wild type does not contain the ``consensus''
binding sequences: $\textrm{TGTGA(N\ensuremath{)_{6}}TCACA}$ for CRP
\cite{Berg1988} and $\textrm{TTGACA(N\ensuremath{)_{18}}TATAAT}$ for
RNAP \cite{Harley1987}, but the wild type is only four mutations
away. Four of the large positive coefficients in
Fig.~\ref{fig:linearcoeff} (positions -54, -34, -9, -8, red circles)
correspond to the mutations needed to get the consensus sequences. 

These inferred coefficients may be compared to the energy matrices
derived from the same data with information theoretic techniques by
Kinney et al.~\cite{Kinney2010a}. There the energy matrices were
inferred separately for CRP and RNAP, and also over many different
experiments, while our regression coefficients were inferred from the
whole sequence data. Correlation between our $\beta$'s and the energy
matrices ranged from 89\%-91\% for CRP binding sites. This is
comparable to the 95\% correlation among energy matrices estimated
from different subsets of the data in \cite{Kinney2010a}. Such an
agreement between a manifestly simple linear-nonlinear model and the
results of a computationally complex optimization of
information-theoretic quantities is truly surprising and encouraging.

Since correlations among various energy matrices for the RNAP binding
are somewhat lower (92\%) \cite{Kinney2010a}, we expect the agreement
between the regression and the information-theoretic methods to be
worse for this case. Indeed, the correlations between $\beta$'s and
energy matrices range between 46\% and 54\%. We expect that this
reduction can be attributed partially to the fact that the energy
matrices were inferred by Kinney et al.\ for CRP and RNAP separately
or jointly in a {\em thermodynamic} model, which assumed a direct
relation between RNAP binding and the transcription rate. It has been
discussed and measured repeatedly \cite{Wall:2009co,Garcia:2012gs}
that transcription rate is strongly affected by kinetics of
transcriptional initiation, which is not modeled for by the
thermodynamic probability of finding RNAP bound to the regulatory
sequence. Unlike the energy matrices, our statistical model inferred
from the entire sequence can account for these kinetic effects, and
may be more accurate in this context.  Since such effects are absent
for transcription factor binding, they can potentially explain the
differences in agreements between the models observed for CRP and RNAP
binding sites. Such kinetic effects may also explain the difference
between the wild type and the consensus (that is, the strongest)
binding sequences mentioned above. Additional biophysical experiments
are needed to carefully explore these issues.

\subsection{Inferring epistatic contributions to fitness}

The simplest model with epistatic interactions between all pairs of
nucleotides is a quadratic or bilinear model, written as:
\begin{equation}
f(y)=\beta_{0}+\sum_{j}\beta_{j}x_{j}+\sum_{i<j}\beta_{ij}x_{i}x_{j}+\varepsilon.\label{eq:quadratic}
\end{equation}
The last sum is over all nucleotide pairs. Here nonzero $\beta_{ij}$
would indicate the presence of pairwise epistasis. For example,
$\beta_i,\beta_j$, and $\beta_{ij}$ all of the same sign is comonly
referred as synergistic epistasis, where contribution of the pair of
mutations is stronger than of each mutation alone. Other possible
types of epistasis are described below.

Note that, in Eq.~(\ref{eq:quadratic}), we keep $f(y)$ the same as in
the previous section, which maximizes the explanatory power of the
non-epistatic terms and minimizes that for the epistatic terms. The
number of epistatic terms in this statistical model ($\sim L^{2}$)
should be contrasted with typical biophysical models of protein-DNA
interactions, which include only a single free energy term describing
interactions between the CRP and RNAP proteins
\cite{Kuhlman2007,Kinney2010a}.

The total number of coefficients $\beta_{0}$, $\beta_{i}$, and
$\beta_{ij}$ in the quadratic epistasis model,
Eq.\,(\ref{eq:quadratic}), is 25,201 (accounting for the fact that, in
a single genome, only one mutation per site is allowed). Overfitting
is a concern since the number of observations, 129,000, is not much
larger than the number of coefficients. To infer a model that does not
overfit, we applied a standard regularization procedure, which
penalizes overly complex models and imposes sparsity on the number of
nonzero interaction terms (see {\em Methods}). Since available
genotypes were not uniformly distributed, but rather biased towards
the wild type, we supplemented traditional cross-validation approaches
with additional checks to ensure that the regularization selects the model with the
highest explanatory power, but no overfitting. The chosen model and
its coefficients are discussed in the following. As we show in {\em
  Methods}, Fig.~\ref{fig:sensitivity}, the general structure of the
inferred epistatic coefficients $\beta_{ij}$ is only weakly dependent
on the specifics of the model choice.

\begin{figure}
\begin{centering}
\includegraphics[width=3in]{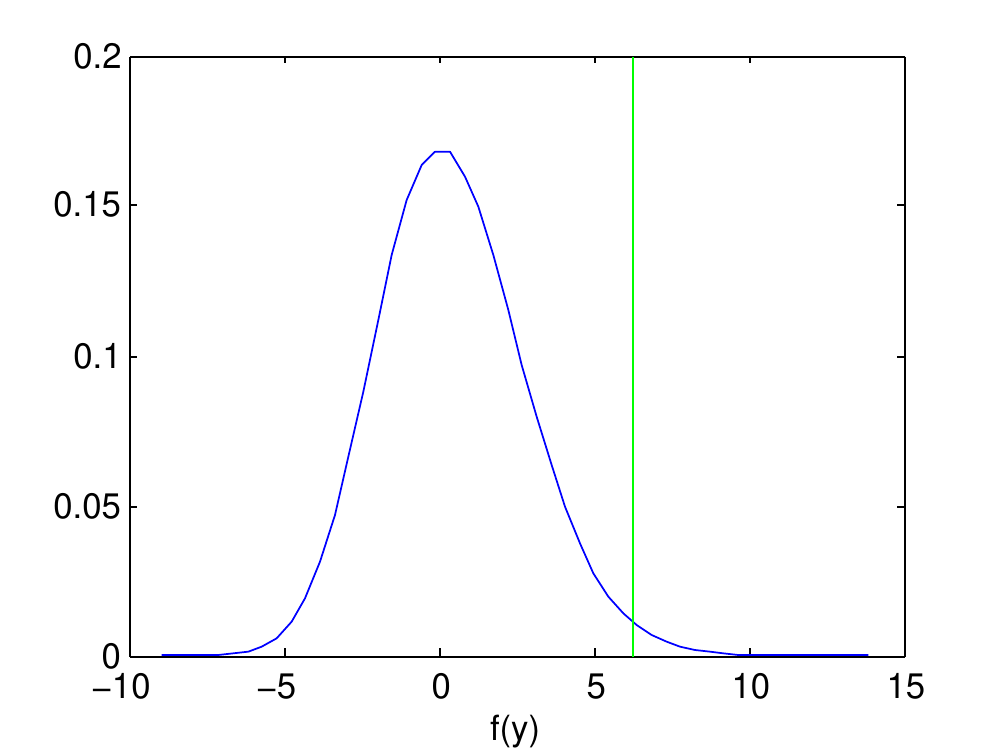}
\par\end{centering}
\caption{\label{fig:opthistrand2}Histogram of phenotype $f(y)$ values of
$10^{5}$ uniformly random sequences for the inferred epistatic model.
Random sequences have very low inferred phenotype values because of the specificity
of binding sites. The peak of the distribution indicates what phenotype
values evolve under neutral conditions. The the wild-type value, $\beta_{0}$
(green line), is much higher than the neutral value indicating selective
pressure.}
\end{figure}

The distribution of inferred phenotype values for randomly generated
sequences (Fig.\,\ref{fig:opthistrand2}) shows that the random
sequences are typically not very functional (presumably because the
binding sites loose specificity). The peak near $f(y)=0$ represents
the most common sequence that would be observed under neutral
evolution, and the relatively high value for the wild-type ($f_{{\rm
    wt}}=6.2$) compared to the random sequences indicates that it is
under strong selection. Notice that we can assert this without any
comparative genomics or population genetics data, which would
typically be required.

The fraction of variance explained by the pairwise epistatic model is
$r_{\textrm{CV}}^{2}=0.571\pm0.007$ (although it is sensitive to the
regularization parameter, cf.~Fig.~\ref{fig:lasso1}). Comparing to the
non-epistatic model with $r^{2}=0.514$, and taking into account the
intrinsic experimental noise of 10-24\%, we see that about 7\% of the
explainable variance is due to the pairwise epistasis. However, it is
possible that more data would increase the amount of predictive power
of the epistatic contributions. Furthermore, combinations of multiple
epistatic interactions may have a net nonepistatic contribution to the
phenotype (but not the other way around). Thus this 7\% figure is, in
many respects, a negatively biased estimate of importance of
epistasis.

The non-epistatic coefficients are about 70\% non-zero, but the
interaction terms are very sparse, about 3\% non-zero. The phenotype
is affected by mutations in some positions more than
others. Coefficients with the largest magnitudes belong to positions
within the CRP and RNAP binding sites (see
Fig.\,\ref{fig:interactions}).  Thus this kind of data allows for
identification of binding sites without a biophysical model of
protein-DNA interactions, as is done traditionally
\cite{Berg1987,Djordjevic2003}. More importantly, as
Fig.~\ref{fig:interactions} shows, the model can infer functional
interactions between amino acid or nucleic acid binding over a much
longer range than can be computed from biophysical and structural
biology approaches \cite{Bauer2010}. The consistency of our results
with known binding sites validates our inferences. Alternative methods
that instead limit the number of inferred coefficients by constraining
the range of interactions, or by allowing interactions only between
consensus sites, would either miss the long-range effects, or the
small (but statistically significant) interactions away from the
binding sites seen in Fig.~\ref{fig:interactions}.

\begin{figure}
\begin{centering}
\raisebox{19\height}{\bf(a)}\includegraphics[height=2.4in]{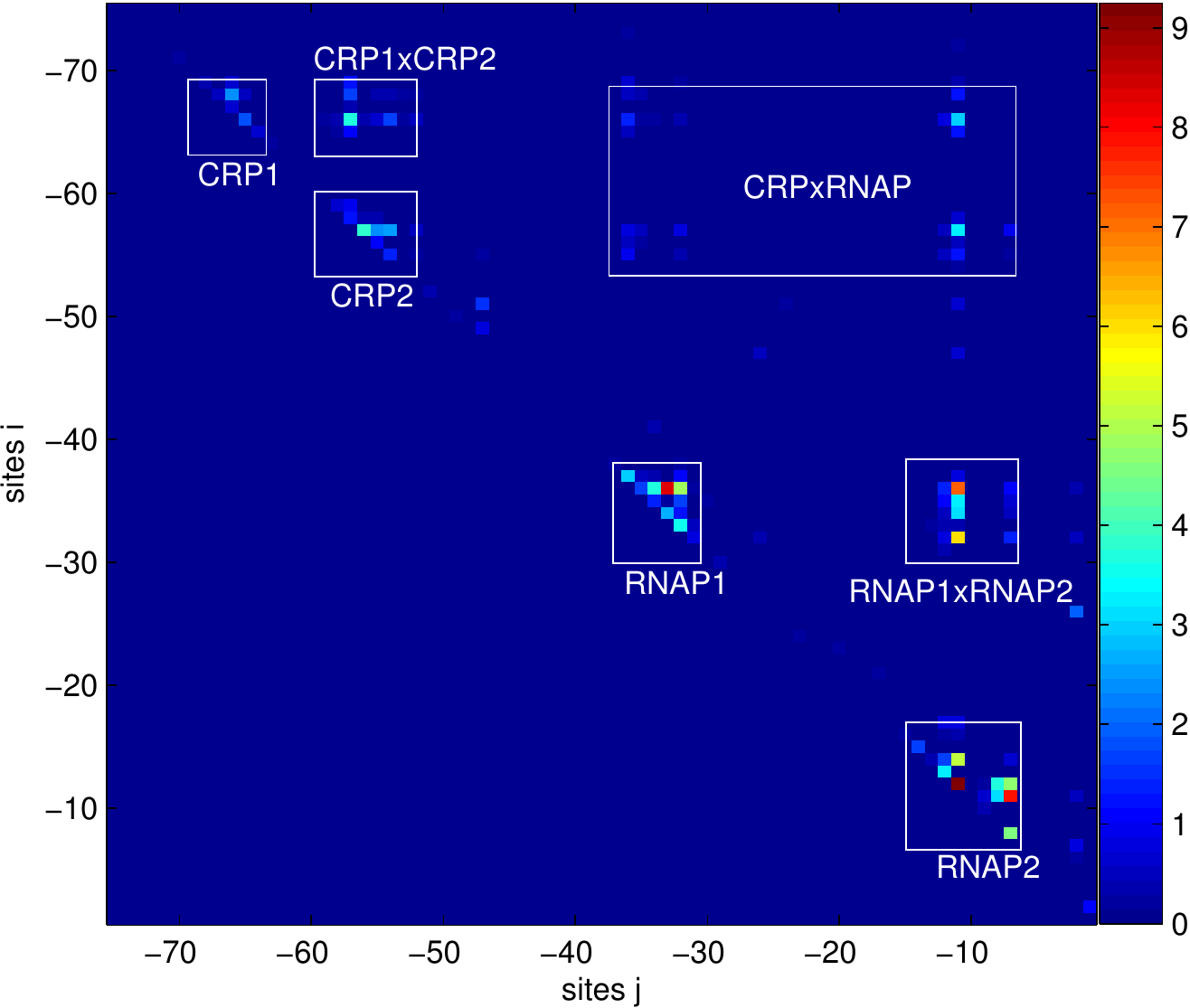}\quad\raisebox{19\height}{ \bf(b)}\includegraphics[height=2.4in]{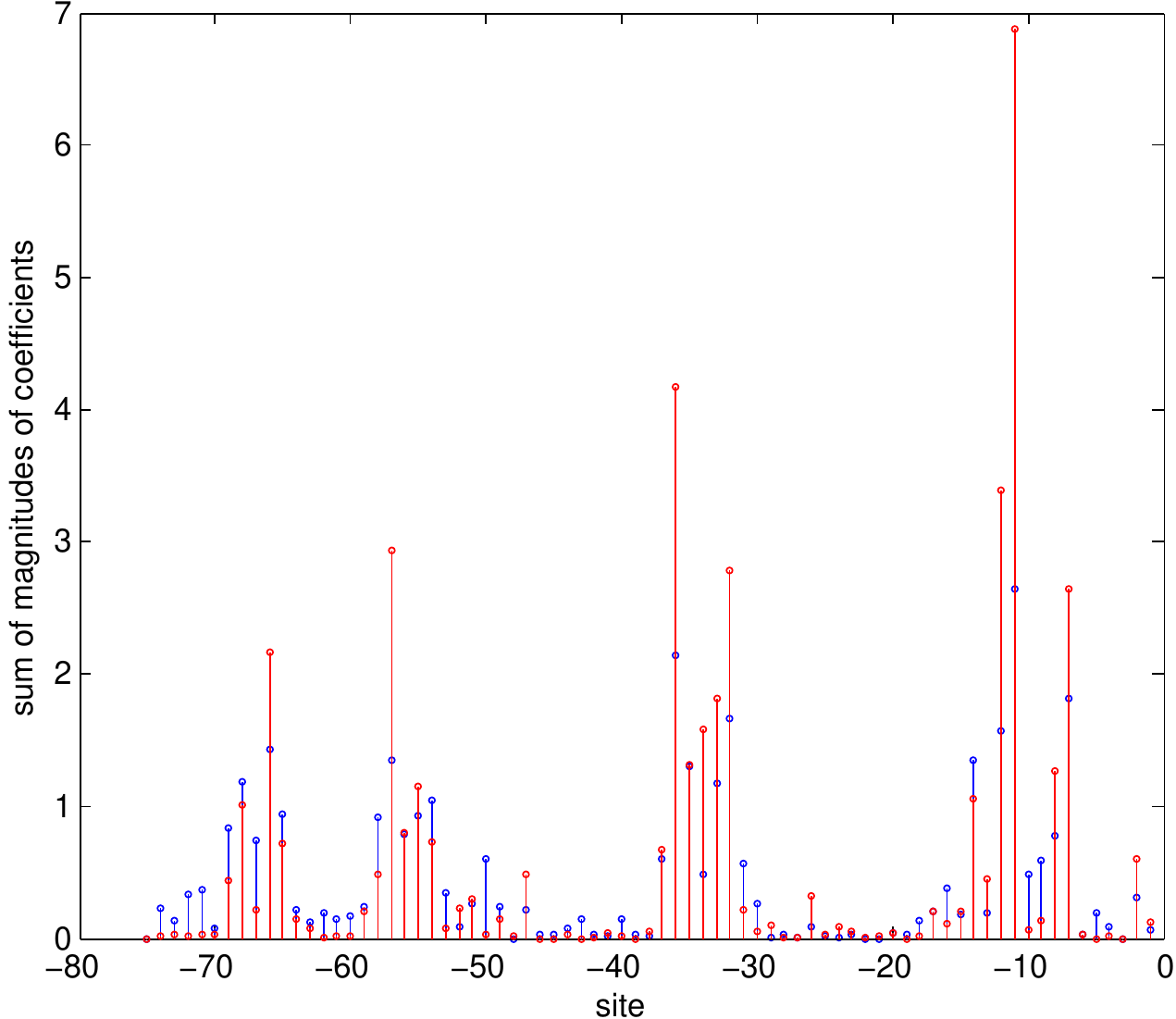}
\caption{\label{fig:interactions} a) Matrix of the sum of the absolute
  values of the pair interaction coefficients for each pair of sites
  $i,j$ (3 mutations per site equals 9 interactions) for the chosen
  statistical model. The clusters
  near the diagonal are interactions within the RNAP and CRP binding
  sites, and the off-diagonal clusters are interactions between the
  binding sites. b) Red: Site-specific sum of absolute
  values of additive coefficients, divided by 3 (the number of
  possible mutations). Black: site-specific sum of absolute values of
  epistatic coefficients, divided by 9 (the number of possible
  mutation pairs). Epistatic and additive effects are strongly
  correlated, with the correlation coefficient 0.90.  }
\par\end{centering}
\end{figure}

The interaction coefficients are observed to be clustered around the
subunits of the system CRP, RNAP, and their constituent binding
sites. The inter- and intra- binding site interactions are easy to
separate in Fig.~\ref{fig:interactions}, allowing a comparison of the
magnitude of the interactions between the subunits, summarized in
Tbl.~\ref{fig:epistasis1}. Interestingly, CRP and RNAP interact on the
same order of magnitude as their constituent binding sites interact
among and within themselves.
\begin{table}
\begin{centering}
\begin{tabular}{|c|c|c|c|c|c|}
\cline{2-6} 
\multicolumn{1}{c|}{} & $\sum|\beta_{ij}|$ & non-zero  & antagonistic & synergistic & sign\tabularnewline
\hline 
all & 194 & 629 & 388 & 56 & 185\tabularnewline
\hline 
CRP1 & 8.2 & 43 & 36 & 1 & 6\tabularnewline
\hline 
CRP2 & 16.1 & 58 & 26 & 5 & 27\tabularnewline
\hline 
CRP1 x CRP2 & 14.5 & 77 & 54 & 4 & 19\tabularnewline
\hline 
RNAP1 & 36.8 & 75 & 58 & 5 & 12\tabularnewline
\hline 
RNAP2 & 49.7 & 88 & 31 & 1 & 56\tabularnewline
\hline 
RNAP1 x RNAP2 & 29.8 & 82 & 64 & 9 & 9\tabularnewline
\hline 
CRP x RNAP & 25.4 & 128 & 115 & 1 & 12\tabularnewline
\hline 
\end{tabular}
\par\end{centering}
\caption{\label{fig:epistasis1} The interaction coefficients for $\lambda=0.021$ are
  clustered around the subunits of the system: CRP, RNAP, and their
  constituent binding sites (defined by white rectangles in figure \ref{fig:interactions}a). The total amount of interaction (sum of the magnitude of coefficients) is shown in the first column. The interactions are
  categorized into three exclusive types of epistasis: synergistic,
  $\beta_{ij}$, $\beta_{i}$, and $\beta_{j}$ share the same sign (and
  are non-zero), antagonistic, $\beta_{i}$ and $\beta_{j}$ share the
  same sign, but $\beta_{ij}$ has opposite sign, and sign epistasis,
  $\beta_{i}$, and $\beta_{j}$ are of opposite sign and $\beta_{ij}$
  is non zero.}
\end{table}

Epistatic interactions may be classified into several categories (see
Table~\ref{fig:epistasis1}): synergistic epistasis (the effect of two
same-sign mutations is larger than the sum of the effects of each one
separately), antagonistic epistasis (the effect of two same-sign
mutations is smaller than the sum of their individual effects), and
other epistatic effects (the individual effects of two mutations have
opposite signs, while epistasis is present). We find that most of the
interactions in the \emph{E. coli lac} promoter are antagonistic
(388/629=62\%). This is likely because mutations change protein-DNA
binding affinity nearly additively, which leads to ``diminishing
returns'' from contributions of individual mutations to
transcriptional activity, similar to \cite{Chou2011,Khan2011a}. Indeed, if the
transcription rate is given by a sigmoidal function of the binding
free energy $F$, such as $\sim 1/(1+e^{F/kT})$ or similar
\cite{Kinney2010a}, then improvements in $F$ are incrementally less
important when it is already large and negative. Thus the effect of
matching an appropriate nucleotide to the corresponding amino acid
decreases when other bases are already matched. Epistasis produced by
this mechanism should be antagonistic, but mild
\cite{Chou2011,Khan2011a}. Indeed, we found only one case of a severe type of
antagonistic epistasis (reciprocal sign epistasis), where the
individual effects are both harmful, but the total effect is
beneficial. It is known that reciprocal sign epistasis is a necessary
(but insufficient) condition for a multi-peaked landscape
\cite{Poelwijk2011a}, and hence we expect this landscape to be fairly
smooth (at most two maxima).

While the relationship between phenotype (transcription) and fitness
is not precisely known in this experiment, they are likely to be
correlated. Therefore the roughness in the genotype-phenotype map is
likely to be important for the whole fitness landscape. Identifying
fitness with $f$, we characterized this roughness by directly
exploring the accessibility of the local optima of the inferred map.
We used an adaptive walk similar to the evolution of a large
population in the weak mutation regime, which can move only towards
higher values and cannot escape local maxima. Starting from the
wild-type sequence, the algorithm only chooses mutations that increase
the phenotype (or fitness), with probability proportional to the log
fitness difference. Out of 1000 random walks, the population ends up
in only two very similar sequences which differ by 2 mutations, and
they are 40 and 39 mutations away from the wild type (compare to the
average of $\sim 6.8$ mutations per sequence). Since the sequences are
so far away from the training data, their predicted phenotype value
are not accurate predictions of the real local maxima.

\subsection{Second and higher order epistasis for a subsequence}
We have insufficient data to study third and higher order epistasis on
the entire 75 bp sequence. However, since most of the linear and the
2nd order epistatic effects in our analysis are concentrated at the
consensus binding sites (cf.~Fig.~\ref{fig:interactions}), we have
performed 3rd order order epistatic analysis on 22 base pairs
subsequences of the data, limited to the four known binding sites in
the sequence. That is, in addition to the linear and the bi-linear
model, we also fitted:
\begin{eqnarray}
  f(y)&=&\beta_{0}+\sum_{j}\beta_{j}x_{j}+\sum_{i<j}\beta_{ij}x_{i}x_{j}+\sum_{i<j<k}\beta_{ijk}x_{i}x_{j}x_{k}+\varepsilon,
\label{eq:cubic}
\end{eqnarray}
where the same procedure was used to find the non-linear function,
$f(y)$ (see {\em Methods}).  Note that the 22 base pairs were selected
based upon consensus binding site locations, not upon our analysis
in the preceding sections. Thus one does not expect overfitting that
would ensue if the same data were used to identify the binding sites
first, and then to refine their epistatic model.

For this subset of nucleotides, the model with only additive effects,
Eq.~(\ref{eq:linear}), had an $r^2=0.41$. The 2nd order epistatic
model, Eq.~(\ref{eq:quadratic}) had $r^2=0.55$. Here the number of
interaction coefficients was much smaller (2,212), resulting in no
signs of overfitting even without regularization. Thus the importance
of quadratic epistasis, which explains 14-20\% of the explainable
variance for the subsequence, is no longer data limited. Like for the
full sequence, we investigated the roughness of the landscape created
by the binding sites subsequence. We found the landcape to be smooth,
with only one global maximum, exactly matching the consensus (but not
the wild type) regulatory sequence.

The 3rd order epistatic model, Eq.~(\ref{eq:cubic}), had 47,972
coefficients, which needed to be regularized in the same way as the
quadratic model ({\em Methods}). This yielded $r^2=0.54$ at maximum
cross validated $r^2$. Thus the higher order interactions do not
improve the fit, and there is {\em no evidence} for these 3rd order
epistatic interactions in the data, although it is possible that
larger data sets would reveal them. Similarly, further restricting the
subset of base pairs used in the analysis did not discover
statistically significant 3rd order effects. In other words, quite
surprisingly, for these data, combinatorial {\em effects of triple
  mutations can be fully modeled by effects produced by constitutive
  pairs of the triples}.

\subsection{Landscape in two environments}

In addition to the data from the three experiments analyzed above,
Kinney et al.~\cite{Kinney2010a} performed experiments with a
different strain of bacteria (TK310) that is unable to control its
intracellular cAMP levels. Because CRP is activated by cAMP, varying
extracellular cAMP levels controls the active intracellular
concentration of CRP. \emph{E.  coli} prefers to metabolize glucose
over lactose, so cAMP is inhibited by the presence of glucose, and
\emph{lac} expression is suppressed when glucose is present. We
inferred genotype-phenotype maps using the non-epistatic model as in
the Section 2.1 for two conditions, no cAMP and $500\mu M$ cAMP,
representing an environment with glucose and no glucose. The datasets
are smaller ($\sim25,000$ sequences), and distinguish only 5 levels of
fluorescence, but they are otherwise very similar, so the same
linear-nonlinear $r^{2}$ optimization was used. The results shown
below were found with the non-epistatic model. However, here the pair
interactions account for a smaller fraction of the variance, and the
epistatic model produces very similar fitted values.

As expected, when CRP is not active there is little binding at the CRP
sites, and the associated coefficients are almost all small
(Fig.\,\ref{fig:campcoeffs}).  Because of the lack of CRP binding,
expression for the wild type sequence, and sequences close to the
wild-type, is lower when there is glucose
(Fig.\,\ref{fig:campcamp}). However, there are some changes to the
RNAP binding site coefficients. Random sequences are not functional in
the no-glucose environment, but they have some small functionality,
comparable to the wild-type, in the glucose environment
(Fig.\,\ref{fig:campcamp}), suggesting that there is less specificity
in the RNAP binding. Note also that some of the coefficients,
especially for the no cAMP case, are large just outside the
traditional RNAP binding domain. Unexpectedly, for no cAMP, the
transcription rate is comparable to the cAMP present case, when CRP
helps polymerase recruitment. This suggests some
additional biophysical binding mechanisms, currently unexplored. As
discussed above, these mechanisms are quite possibly kinetic in
nature.

In the no cAMP (glucose) environment, \emph{lac} expression should
decrease the growth rate because the cell is metabolizing glucose
instead of lactose, and \emph{lac }expression costs resources
\cite{Dekel2005,Perfeito2011}. Therefore we expect sequences under
selection, such as the wild type, to have relatively high expression
with cAMP, and low expression without cAMP, compared to sequences not
under selection (random sequences). Figure \ref{fig:campcamp} shows
that there exist very few sequences which are better than the wild
type in both environments, i.e. simultaneously higher expression with
cAMP, and lower expression without cAMP. The non-elliptical shape of
the fitted values for the experimental sequences suggests again that
the wild type is under a strong selection towards the top left corner
of the plot. Finally, we point out that, even when lactose is being
metabolized, too high expression of {\em lac} genes is costly,
possibly because cellular resources are pulled to {\em lac}
transcription and translation and away from production of essential
proteins \cite{Dekel2005}. This may make sequences in the top right
corner of Fig.~\ref{fig:campcamp} less fit than our monotonically
increasing $f(y)$ model assumes, making the wild type even closer to
the global optimality.

\begin{figure}
\begin{centering}
\includegraphics[width=3.5in]{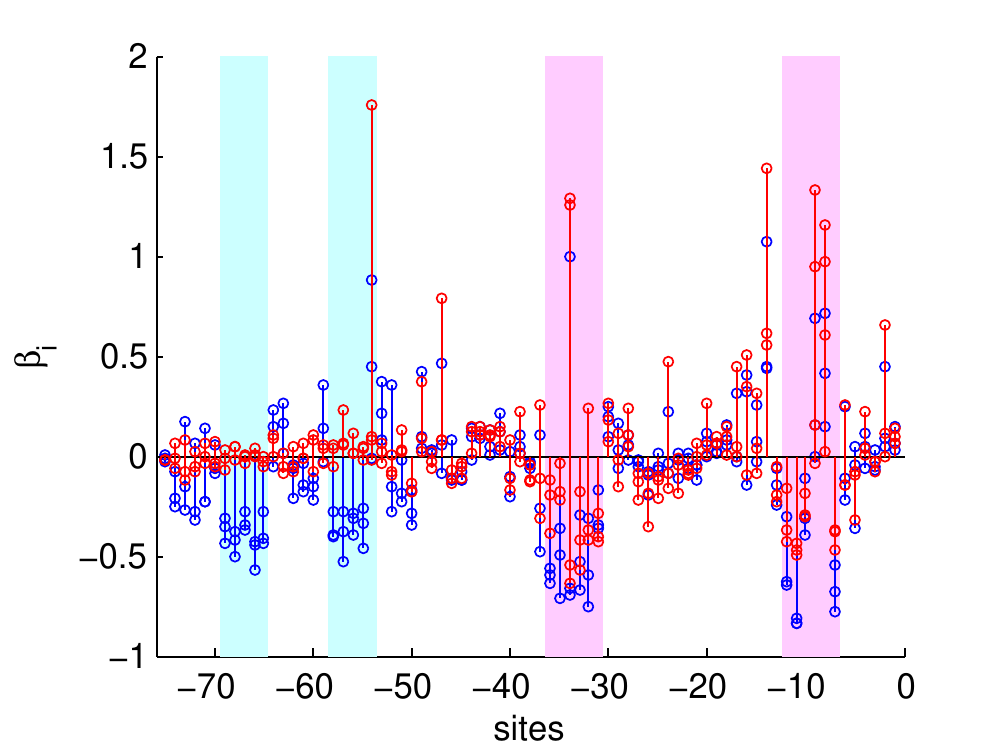}
\par\end{centering}
\caption{\label{fig:campcoeffs}(blue) coefficients $\beta_{i}$ for the non-epistatic
model with no-glucose (normal levels of cAMP) (red) with glucose (no
cAMP). CRP is activated by cAMP and does not bind without it.}
\end{figure}
\begin{figure}
\begin{centering}
\includegraphics[height=3in]{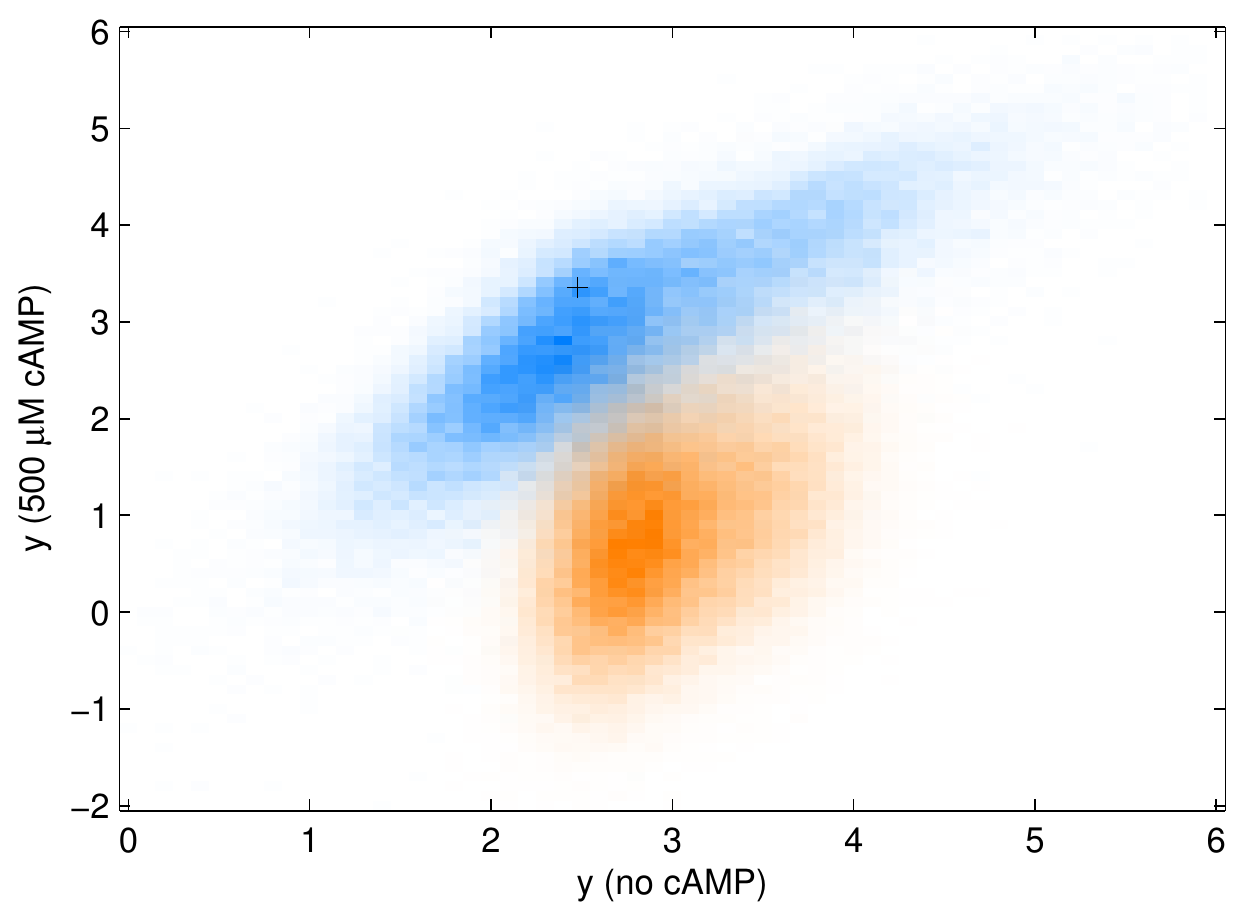}
\par\end{centering}
\caption{\label{fig:campcamp}2D histogram of expression for the two
  environments, no cAMP (glucose), and cAMP (no glucose) for $10^{5}$
  random sequences (orange), and sequences from the experiment (blue),
  which are closer to the wild type (plus sign). The wild-type is
  nearly on the optimal front in that very few sequences have both
  higher expression with cAMP and lower expression without cAMP (above
  and to the left of the plus sign).  The phenotype values range from
  1 to 5 in these experiments. The dis-similarity of measured
  expression and expressions predicted for random sequences along the
  vertical, but not the horizontal axis, likely signals presence of
  poorly understood biophysical mechanisms differentially employed in
  the two considered environments.}
\end{figure}

\section{Discussion}

We constructed a genotype-to-phenotype mapping, including effects of
all pairwise and some higher order epistatic interactions. This was
done by analyzing functional properties of over $100,000$ randomly
mutated sequences in the vicinity of the wild type \emph{E.~coli lac}
operon, queried under different experimental conditions. The control
of dimensionality for the epistatic models, along with the large size
of the dataset, allows for a much more detailed analysis of epistasis
in this bacterial genetic regulatory region.

Our approach is generally similar to those in
Refs.~\cite{Hinkley2011,Kouyos2012}.  However, there are substantial
differences beyond a different model organism used. Our alleles are
nucleotides in a regulatory region of a bacteria, instead of amino
acid variants. Our landscape is more complete, in that interaction
among all pairs of nucleotides in the sequence are estimated from the
data that includes each such pair at least 20 times in different
genetic backgrounds. In particular, we have relaxed the condition
\cite{Mora2010} that the interaction terms $\beta_{ij}$ can depend
only on the distance between the loci, rather than on the specific
positions of the loci.  Mora et al.\ \cite{Mora2010} used maximum
entropy approaches to infer a fitness landscape, while, along with
Hinkley at al.\ \cite{Hinkley2011}, we have focused on linear
regression (though with different regularization constraints and
different nonlinear mapping between the fitness and the observed
phenotype). The epistatic model, Eq.~(\ref{eq:quadratic}), is the same
in the regression and the maximum entropy approach. However, the
philosophical basis behind the approaches is different, and so are the
criteria used to specify the coefficients $\beta$. Maximum entropy
methods choose them to constrain observable correlation functions,
while regression attempts to approximate the entire fitness
function. It remains to be seen which of the two frameworks provides a
better model for genomic data.

Possibly the largest difference from the previous approaches that
considered epistatic interactions for many mutations is that we found
a genotype-phenotype map, rather than the true fitness
landscape. While we expect the phenotype and the fitness to be
strongly correlated when lactose is being metabolized (and
anti-correlated otherwise), the relation between the fitness and
either the observed fluorescence or its nonlinearly reparameterized
form, $f(y)$, is likely nontrivial. Ideally, a second experiment would
measure the phenotype-to-fitness map to complete the reconstruction of
the fitness landscape. In fact, Dekel and Alon\cite{Dekel2005} have
completed this second step for the {\em lac} regulatory
sequence. However, we cannot use their findings since their {\em
  E.~coli} strains and growth environments were slightly different
from those of Kinney et al.~\cite{Kinney2010a}.

Binding energy-fitness maps have been inferred from genome wide
studies of transcription factor binding sites using genomic statistics
and population genetics models
\cite{Gerland2002,Berg2004,Mustonen2005,Mustonen2008a}.  In those
studies, the genotype-phenotype maps were largely assumed to be
non-epistatic, in contrast to our work.  It would be interesting to
combine the methods to make a more complete account of epistasis from
genotype to fitness.

Our observations have revealed a few cautionary notes regarding using
genome frequency in a population to reconstruct fitness landscapes
\cite{Mora2010,Hinkley2011}. In such experiments, all sequence data
(including whatever part of it that is left for cross-validation) are
localized near the wild type, near-optimal sequences due to
selection. Carefully inferred models (whether regression or maximum
entropy based) perform well for the observed data, but will generalize
badly for sequences far away from the wild type. Our approach samples
the genotype space more evenly without selection, and therefore is
better suited for making inferences about the global landscape
properties, such as its ruggedness. Nonetheless, even in our data,
with each sequence $\sim 7$ mutations away from the wild type,
extrapolation to much larger genotypic differences produces absurd
results, even if cross-validation fails to notice problems,
cf.~Fig.~\ref{fig:lasso1}.

In our inferred landscape, epistasis accounted for about 7\% (about
15\% for the binding sites subsequence) of the explainable
variance. Most of the epistasis was antagonistic, but the landscape
was essentially single peaked. This is similar to properties of
epistasis in metabolism \cite{Chou2011,Khan2011a}, and the explanation
for both likely involves diminishing returns from successive
individual mutations. It is useful to contrast these findings with the
work on HIV \cite{Kouyos2012} or protein fitness landscapes
\cite{Weinreich:2006ig}, which have observed more substantial
epistasis and many more local maxima.  While it is possible that more
epistatic effects would be observed for our system if more data
were available, more intriguing is the following observation. During
model selection (see {\em Methods}), it was noticed that, due to most
of the sequences being $<10$ mutations from the wildtype, it was
possible to make large prediction errors for sequences with more
mutations. In other words, there was a large extrapolation error for
sequences outside of the training data, and this led to choosing a
more constrained model for final analysis. A less constrained model
(which maximizes $r_{\rm CV}$ in Fig.~\ref{fig:lasso1}) is much more
epistatic, with adaptive walks indicating many local maxima. The
severity of the problem correlates with the nonuniformity of the
genotype sampling, making the data from populations under strong
selection especially suspect. To allow studying global properties of
landscapes, an ideal experiment would sample the sequence space much
more uniformly to avoid extrapolation.

In addition to the weak epistasis, we also found that the wild-type
{\em E.~coli lac} regulatory region is optimal for the two
environments measured. That is, it is on the front of possible
sequences which maximize expression when it is beneficial, and
minimize expression when it is harmful. If under the growth
conditions the fitness is a non-monotonic function of the
transcriptional activity and decreases at high expression
\cite{Dekel2005}, the wild type operon may be not only nearly multi-objective
optimal, but nearly globally optimal. To investigate this, experiments
are needed that would study fitnesses of many sequences under
selection in fluctuating environments.

The ability of our method to identify protein binding sites and
epistatic interactions among them raises an important point. These
epistatic interactions, inferred by either of the methods we have
mentioned in this work, especially interactions over long ranges, may
not correspond to true biophysical interactions between amino acids
and nucleotides. They are likely {\em effective} interactions
resulting from collective effects of many other epistatic terms,
including higher order terms, or a small number of interactions, such
as binding between CRP and RNAP. While there is an admirable
similarity between our linear regression coefficients and energies of
protein-DNA interactions, our approach may not be as informative where
there is enough information to build a detailed biophysical model, but
there are few places in the genome where this is the case. On the
other hand, our approach can detect long distance epistasis, or
non-thermodynamic effects on transcription where a priori it is
unclear that these effects and interactions exist. When working on the
genome scale, effective models that can make accurate {\em
  predictions} of phenotype or fitness for previously unobserved
sequences may be useful regardless of their lack of microscopic
accuracy.  They may be closer to the right level of description of the
problem \cite{Goldenfeld:1999wv}, by striking a balance between
microscopic biophysically relevant detail, and power to describe the
richness of phenomena emerging on the genomic scale. As an example of
this utility, here we found that, for the 22 bp long subsequence of
the regulatory region that includes the binding sites, there was no
evidence for 3rd order epistatic effects. The fact that pairwise {\em
  effective} interaction models, with only a few higher order
contributions, provide excellent fits to multivariate data has been
observed by now in the context of neurophysiological recordings
\cite{Schneidman2006,Tkacik2006,Tang2008,Cocco2009,Ohiorhenuan2010},
microarray-measured gene expressions
\cite{Margolin2006,Wang2009,Margolin2010}, and sequencing data
\cite{Mora2010}, to which our analysis has just added another
example. These frequent successes of pairwise models in diverse
domains are certainly surprising and, as of now, unexplained. They
raise many interesting questions about general theories of
multivariate biological data, which are still waiting for their
answers.

\ifPNASstyle
   \begin{materials}
\else 
   \section{Methods}
\fi
\subsection{Preparation of the dataset}

\begin{table}[t]
\begin{center}\begin{tabular}{|c|c|c|c|c|}
\hline 
 & A & T & C & G\tabularnewline
\hline 
A &  & 100 & 010 & 001\tabularnewline
\hline 
T & 001 &  & 100 & 010\tabularnewline
\hline 
C & 010 & 001 &  & 100\tabularnewline
\hline 
G & 100 & 010 & 001 & \tabularnewline
\hline 
\end{tabular}\end{center}
  \caption{\label{tbl:code} Mutation encoding scheme (dummy variables). For a wildtype
    nucleic acid (vertical) a mutation to another nucleic acid
    (horizontal) is encoded by the corresponding sequence.}
\end{table}

To make inferences on the largest dataset possible, we combined the
data from three experiments done by Kinney et al.~\cite{Kinney2010a}
(fullwt, crpwt, rnapwt, 129,000 sequences total), which differ only by
the regions in which mutations were allowed to take place. Fullwt was
mutagenized over the whole sequence (-75:-1), while crpwt and rnapwt
were mutagenized only over the CRP binding area and RNAP binding area.
In addition, some sequences were rejected for data quality reasons:
identical sequences in the same bin were likely to be not independent
measurements (see Supplemental Materials in Ref.~\cite{Kinney2010a}),
and sequences with an exceptional number of mutations ($>20$) were
probably errors.  

\subsection{Linear-nonlinear model}

Part of the genotype-phenotype map may be non-linear due to the
mapping from fluorescence to bin number and some remaining background
fluorescence. To identify pairwise interactions in the background of
an arbitrary mean nonlinear genotype-phenotype map, we introduce a
generalized linear-nonlinear model:
\begin{equation}
f(y)=\beta_{0}+\sum_{j}\beta_{j}x_{j}+\varepsilon,
\label{linear}
\end{equation}
where $f(y)$ is a monotonically increasing, nonlinear function of
$y$. The function is found by maximizing the fit ($r^{2}$), which
corresponds to minimizing\footnote{This method resembles a type of
  generalized linear model called ordinal probit
  regression\cite{Green2003}, and is also similar to the inference of
  non-linear filters in computational neuroscience using
  information-theoretic tools \cite{Sharpee2006a}.}
\begin{equation}
f(y)=\arg\min_{g(y)}\frac{\textrm{var}\left(g(y)-\beta_{0}-\sum_{j}\beta_{j}x_{j}\right)}{\textrm{var}\left(g(y)\right)}.
\end{equation}
We add the constraints that $f(9)=9$, and $f(1)=1$ to keep
$\textrm{var}\left(g(y)\right)$ finite. The function $g(y)$ is defined
over only 9 values of $y$, and a constrained non-linear optimization
procedure ({\tt fmincon} from MATLAB) finds an optimal $f(y)$ quickly
(Fig.~\ref{fig:optimaly}). 

The summary statistics change when replacing $y$ with $f(y)$. The
variance of the bin numbers increases from 6.5 to 7.6, and the $r^2$
increases from 0.476 for the linear model for $y$, to 0.514 for the
linear model for $f(y)$. The experimental noise estimates (see below)
are also slightly different.

\begin{figure}
\centerline{\includegraphics[width=3in]{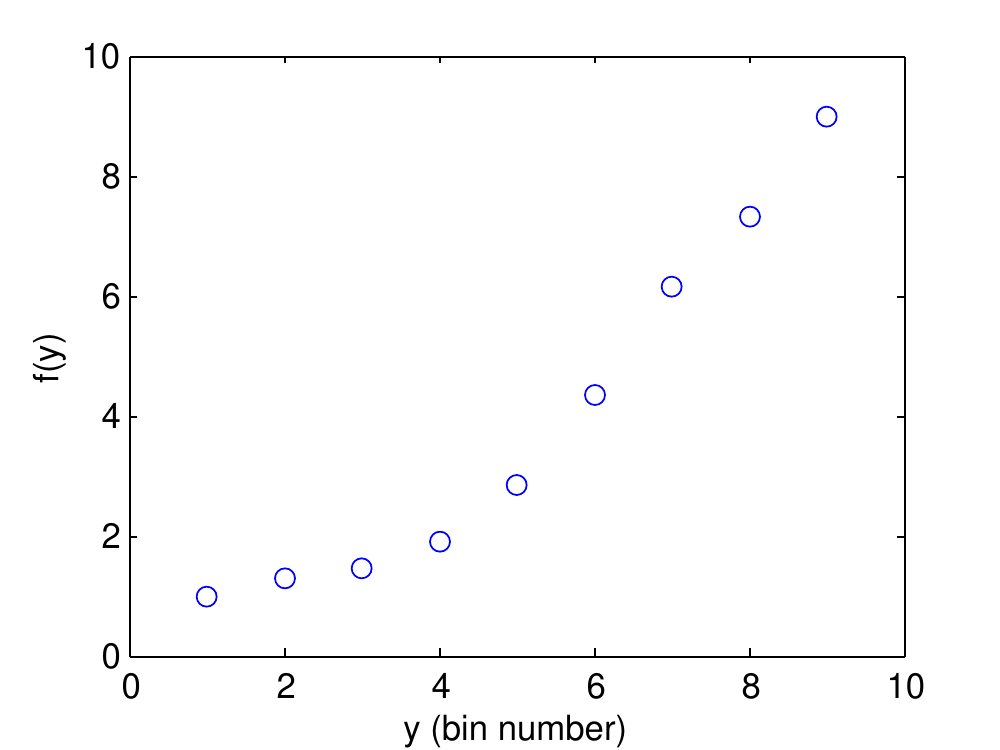}}
\caption{\label{fig:optimaly}Generalizing the fitted function by replacing
the output values $y$ with a non-linear function $f(y)$ improves
the least squares fit. Constrained non-linear optimization found the
optimal $f(y)$ for the linear model with $r_{opt}^{2}=0.514\pm0.002$.
The non-linearity is due to the first few bins being dominated by
background fluorescence and not gene expression.}
\end{figure}

Assuming a monotonic relationship between genotype and phenotype,
$f(y)$ is the function that maximizes the phenotype prediction from
the non-epistatic (linear in $x_{i})$ contributions. This reduces the
amount of variability left to be predicted by {\em any} epistatic
model, whether of genotype-phenotype map, or genotype-fitness map
(provided that the fitness is monotonically related to the
phenotype). This also prevents the epistatic model from fitting any
average non-linear effects. Thus our subsequent assessment of
importance of the epistasis should be viewed as biased towards
underestimation.

\subsection{Estimates of intrinsic noise in the data}
Experimental data is corrupted by errors in both fluorescence
measurements and sequencing. One estimate of this intrinsic noise is
obtained by averaging the variance of $f(y)$ for identical sequences
with different recorded fluorescence values.  The ratio of this
intrinsic variance to the total variance of $f(y)$ is $1.8/7.6=0.24$.
Since this excludes all sequences that fell into just one bin and have
an unknown variance $<1$, this estimate is an upper bound on the noise
variance.

Another estimate can be obtained by using the controls from
Ref.~\cite{Kinney2010a}, which provide fluorescence numbers for many
individual wild type bacteria. The fluorescence variance in optimized
bin units is 0.74, which is $0.74/7.6=0.10$ of the data variance. This
number underestimates the average noise since wild type bacteria
express strongly, so that the fluorescence noise for them is smaller
than for most other sequences.

\subsection{Regularization and model selection} 

Statistical model with the number of parameters comparable to the data
set size may overfit, that is, model statistical noise in the data. To
prevent overfitting, we minimize the mean squared error in
Eq.\,(\ref{eq:quadratic}) subject to a regularizing constraint
\begin{equation}
  \beta^*=\arg \min_\beta
  \left(\langle\varepsilon^{2}\rangle+\lambda\left\Vert
      \beta\right\Vert \right),
\label{LASSO}
\end{equation}
where $\beta$ is the concatenated vector of all the regression
coefficients, $||\beta||$ is its norm, and $\lambda$ is a free
parameter (Lagrange multiplier), unknown {\em a
  priori}. Regularization constrains the statistical complexity of the
model by minimizing the norm of the coefficients
\cite{MacKay2003}. When the $L_1$ norm is used, $\Vert \beta\Vert
=\sum|\beta_{i}|$, this regression is called the Least Absolute
Shrinkage and Selection Operator (LASSO) \cite{Tibshirani1996}. LASSO
favors sparse solutions, which is a reasonable assumption since most
of the $\beta$'s are interaction terms, and interactions are presumed
to be mainly between the relatively small CRP and RNAP binding
sequences. Thanks to an efficient implementation of the algorithm
\cite{Friedman2010}, we can compute the LASSO solution for 100
different values of $\lambda$, from the maximum value (where the
solution is all $\beta$'s equal to zero), to four orders of magnitude
smaller.

\begin{figure}
\centerline{\includegraphics[width=3.5in]{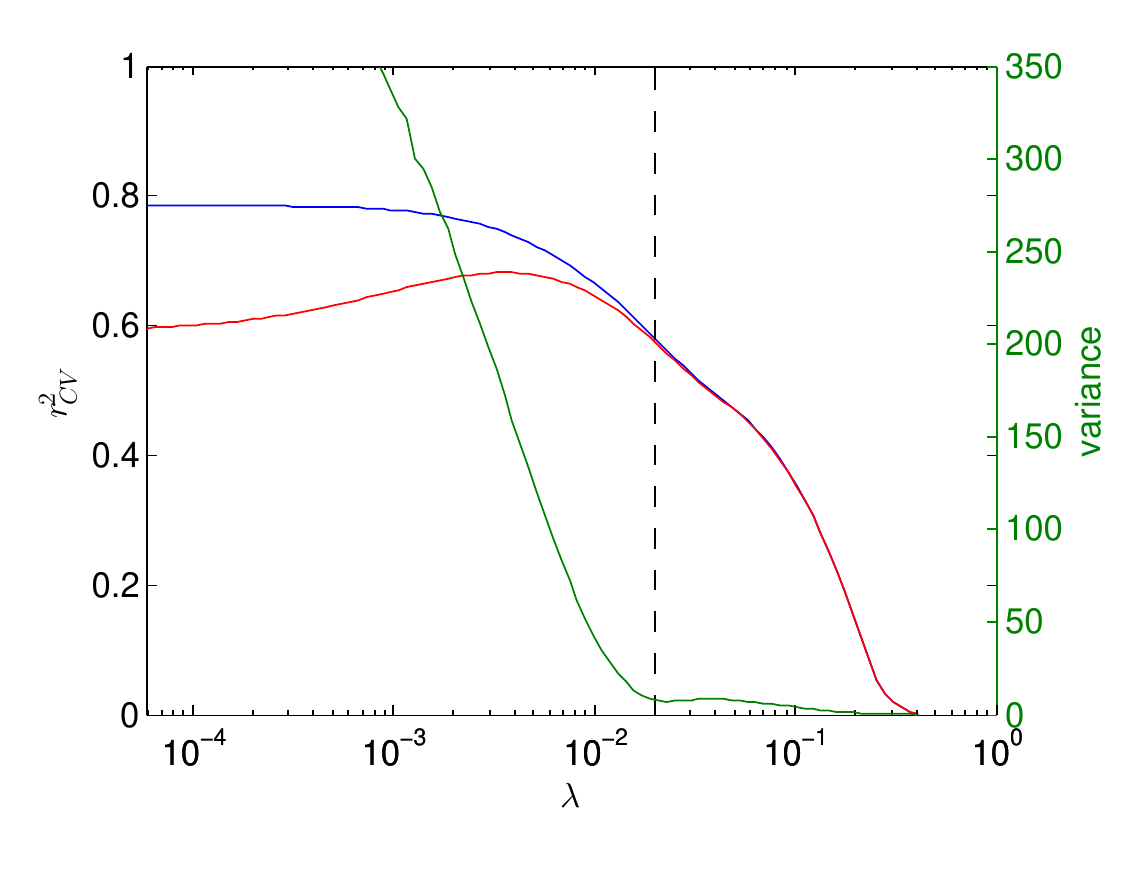}}
\caption{\label{fig:lasso1}The LASSO solution of the quadratic model was computed
for 100 values of $\lambda$. Blue is the $r^{2}$ value, and red
is the 10-fold cross-validated $r_{{\rm CV}}^{2}$. The green curve
is the variance of $f(y)$ for randomly generated sequences. The variance
is too large even for values of $\lambda$ that are larger than the
optimal value predicted by the maximum of the $r_{{\rm CV}}^{2}$
curve. We choose the model with $\lambda=0.021$ (dashed line) for
further analysis. This model has $\sim10^{3}$ non-zero coefficients,
most of which are epistatic.}
\end{figure}

However, choosing the \emph{best} solution (i.e., the right $\lambda$)
is ambiguous. A common method of model selection is cross-validation.
Figure \ref{fig:lasso1} shows that solutions with large $\lambda$ are
a poor fit, while small $\lambda$ values have less predictive power,
as seen through cross-validation. Typically one chooses the best model
as the one with the maximum $r^{2}$ ($r_{\textrm{CV}}^{2}$)
\cite{Tibshirani1996}. However, both the training and the
cross-validation data are sequences with an average of only 6.8
mutations from the wild-type (9\% mutated sites). Thus
cross-validation may not ensure predictability for sequences farther
away in the genotype space. Indeed, the variance of the fitted values
of $f(y)$ for the experimental data is not sensitive to changes in
$\lambda$ (not shown). Nonetheless, Fig.~\ref{fig:lasso1} shows that
the variance of $f(y)$ for random sequences blows up for less
constrained models (low $\lambda$), where unrealistically high fitted
values of $y$ or $f\sim50\dots100$ emerge. This indicates overfitting
due to uneven sampling of the genotype space and the resulting
correlations in the training and the test data. We thus limit
$\lambda$ to the range where the variance of the fitted values for
random sequences is comparable to that for the experimental data and
is insensitive to $\lambda$. Incidentally, this is also the place
where $r^{2}$ and $r_{\textrm{CV}}^{2}$ curves split in Figure
\ref{fig:lasso1} (dashed line, $\lambda=0.021$, 629 non-zero
coefficients). Finally, Fig.~\ref{fig:sensitivity} shows that the general
structure of the solution is only weakly dependent on the exact choice
of $\lambda$.

\begin{figure}
\begin{centering}
\raisebox{19\height}{\bf(a)}\includegraphics[height=2.4in]{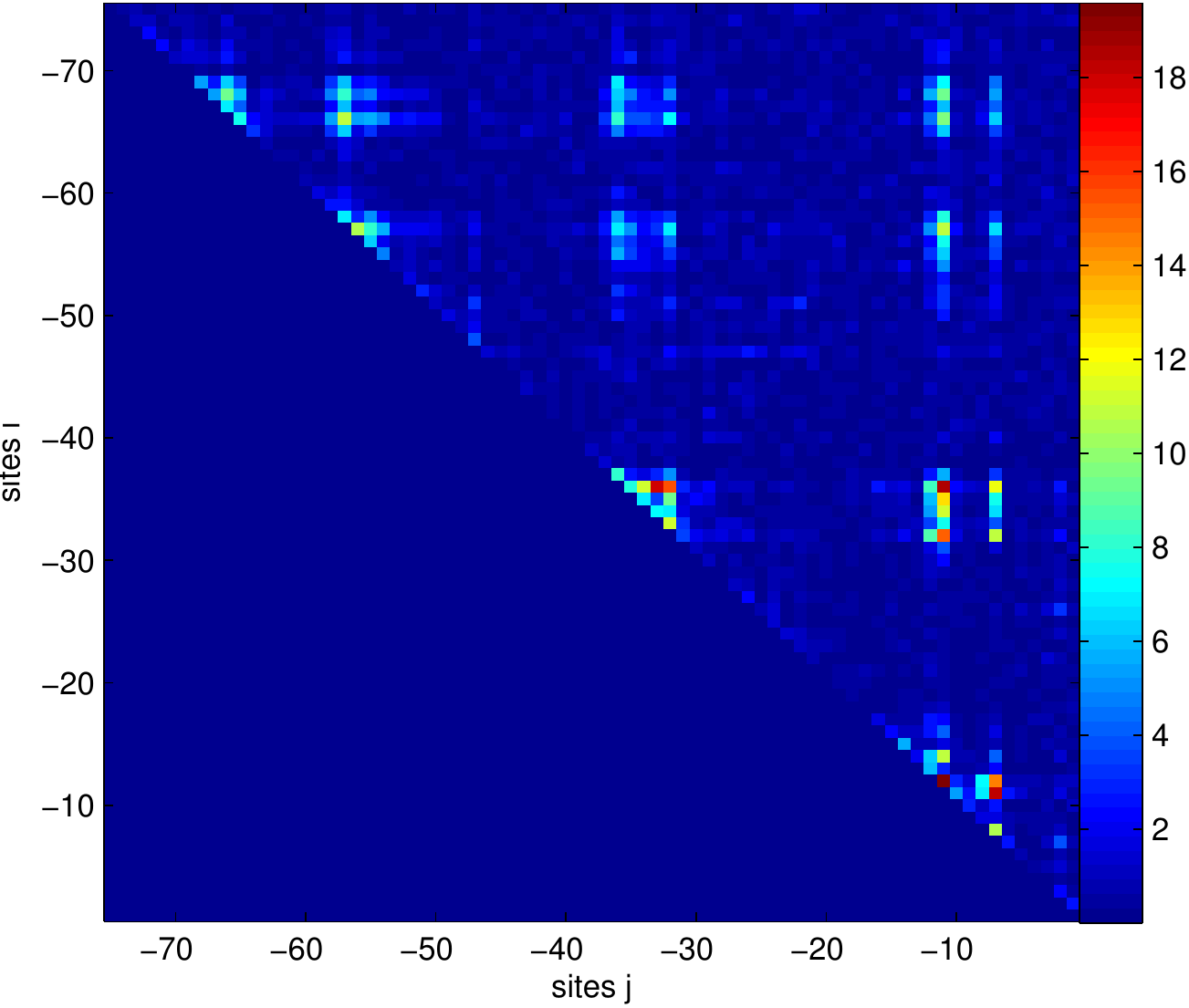}\quad\raisebox{19\height}{ \bf(b)}\includegraphics[height=2.4in]{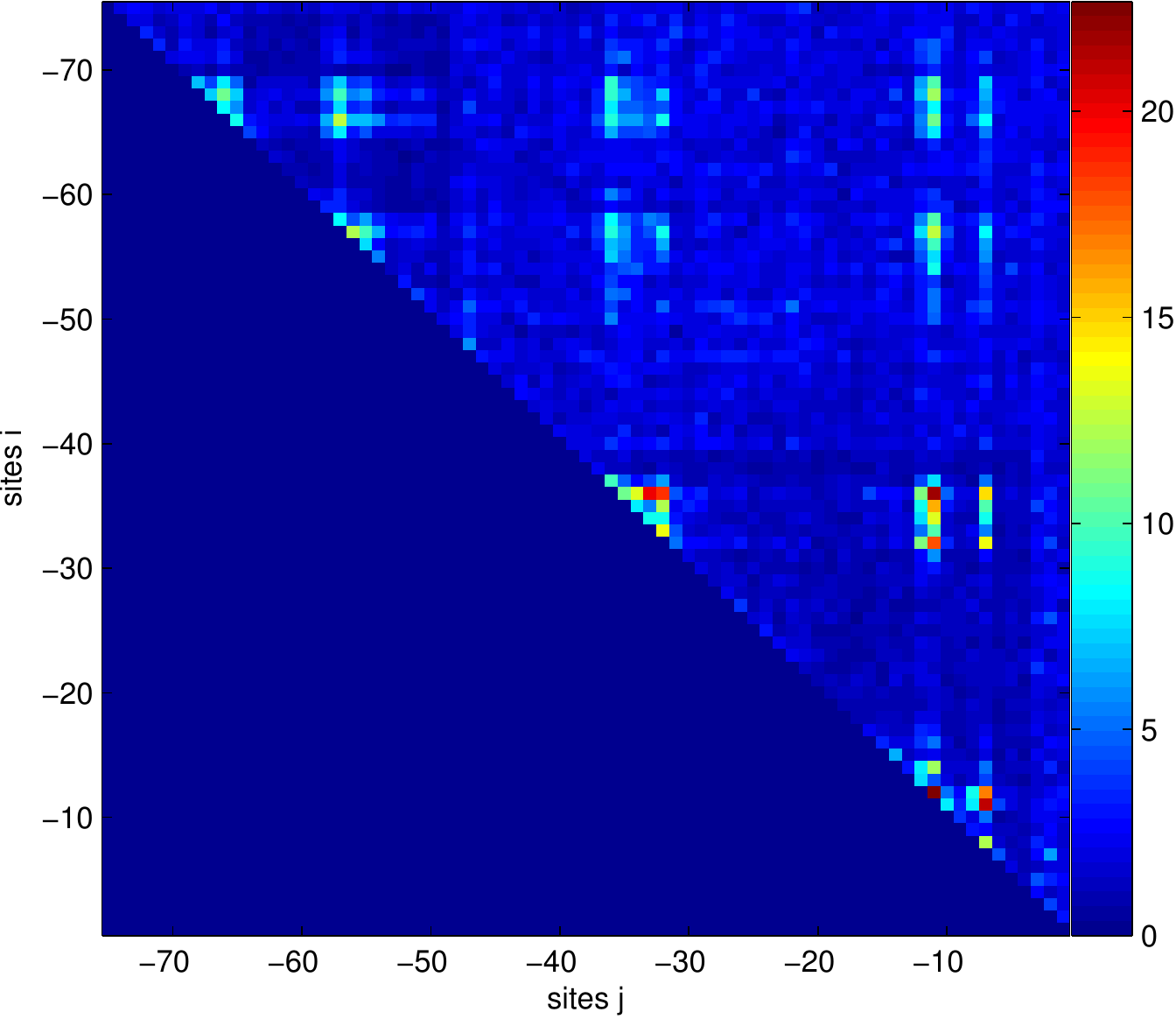}
\caption{\label{fig:sensitivity}Sensitivity of the epistatic
  coefficients to the choice of the regularization parameter
  $\lambda$. As in Fig.~\ref{fig:interactions}, we show the matrices
  of the sums of the absolute values of the pair interaction
  coefficients for each pair of sites $i,j$. a) Coefficients for the
  model with maximum $r_{\textrm{CV}}^{2}$ ($\lambda=0.0032$). b)
  Coefficients for the full model: $\lambda=0$. Notice the same
  general structure of the coefficients for varying $\lambda$,
  including $\lambda=0.021$ in Fig.~\ref{fig:interactions}. This
  indicates stability under changes of the parameter.  }
\par\end{centering}
\end{figure}

\ifPNASstyle
   \end{materials}
   \begin{acknowledgements}
\else
   \section*{Acknowledgements}
\fi

We would like to than Justin Kinney for providing us with the sequence
data, David Cutler, Thierry Mora, and Minsu Kim for illuminating
discussions, Thierry Mora, Justin Kinney, and Philip Johnson for
commenting on the manuscript, and Bruce Levine for general guidance.
\ifPNASstyle
   \end{acknowledgements}
\fi


\ifPNASstyle
   \end{article}
\fi

\end{document}